\documentclass[aps,prd,twocolumn,groupedaddress,nofootinbib]{revtex4}
\usepackage{amsmath}
\usepackage{bm}
\usepackage{epsfig}
\usepackage{graphicx}

\begin{document}

\preprint{ANL-HEP-PR-03-065}

\title{Inclusive quarkonium production and the NRQCD factorization 
approach}


\author{Geoffrey T. Bodwin}
\thanks{Work in the High Energy Physics Division at Argonne National
Laboratory is supported by the U.~S.~Department of Energy, Division of
High Energy Physics, under Contract No.~W-31-109-ENG-38.}
\affiliation{
HEP Division,
Argonne National Laboratory, 9700 South Cass Avenue, Argonne, IL 60439
}


\date{\today}

\begin{abstract}
I discuss the current status of the comparison between quarkonium-production 
measurements and the predictions of the NRQCD factorization formalism.
\end{abstract}


\maketitle

\section{NRQCD factorization}

In calculating rates for quarkonium production or decay, one would like
to separate the short-distance heavy-quark-antiquark ($Q\overline Q$)
annihilation or production process, which has a typical momentum scale
$p\agt m_Q$ and can be treated perturbatively, from the long-distance
quarkonium dynamics, which have a typical momentum scale $p\alt m_Qv$ and
are nonperturbative in character. Here $m_Q$ is the heavy-quark mass, and
$v$ is the relative velocity of the $Q$ and $\overline Q$ in the
quarkonium rest frame. Such a separation of long- and short-distance
scales in heavy-quarkonium production and decay can be expressed
elegantly in terms of the effective field theory Nonrelativistic QCD
(NRQCD) \cite{BBL}.

In the cases of inclusive quarkonium production at large transverse
momentum $p_T$ and at large CM-frame momentum
$p^*$, the cross section can be written in a factorized form as a sum of
products of NRQCD matrix elements and short-distance coefficients:
\begin{equation}
\sigma(H)=\sum_n \frac{F_n(\Lambda)}{m_Q^{d_n-4}}\langle 0|
{\cal O}_n^H(\Lambda)|0\rangle.
\label{prod-fact}
\end{equation}
The $F_n(\Lambda)$ are short-distance coefficients. They are,
essentially, the partonic cross sections to make a $Q\overline Q$ pair
convolved with parton distributions. The $F_n(\Lambda)$ can be
calculated as expansions in the strong-coupling constant $\alpha_s$. 
The (vacuum) matrix elements involve four-fermion operators,
which have the form
\begin{equation}
{\cal O}_n^H=\chi^\dagger \kappa_n\psi
\biggl(\sum_X |H+X\rangle\langle H+X|\biggr) \psi^\dagger
\kappa'_n\chi.
\end{equation}
Here,  $\psi$ is the Pauli spinor field that annihilates a heavy quark,
$\chi$ is the Pauli spinor field that creates a heavy antiquark, and
$\kappa$ contains Pauli matrices, color matrices, and covariant
derivatives. The operator creates a $Q\overline Q$ pair in a state with
certain color, spin, and orbital-angular-momentum quantum numbers,
projects it onto an intermediate state containing a heavy quarkonium $H$
plus anything, and annihilates a $Q\overline Q$ pair with specific quantum
numbers from that state. The operator matrix elements contain all of the
long-distance (nonperturbative) physics. They are the probabilities for
a $Q\overline Q$ pair to evolve into a heavy quarkonium. 

A similar factorization formula applies to inclusive quarkonium decays.
The decay matrix elements are the crossed versions of quarkonium
production matrix elements. Only the color-singlet production
and decay matrix elements are simply related.

The NRQCD factorization formalism gains much of its predictive power
from the fact that the nonperturbative operator matrix elements are
universal, {\it i.e.}, process independent. Although some decay matrix
elements have been computed on the lattice
\cite{Bodwin:1996tg,Bodwin:2001mk}, in general, the matrix elements must
be extracted phenomenologically. The consistency of the phenomenological
matrix elements from process to process is a key test of the NRQCD
factorization formalism.

NRQCD also predicts velocity-scaling rules \cite{BBL}, which give the
leading power behavior of the matrix elements as functions of $v$. It
follows that the sum over operator matrix elements in
Eq.~(\ref{prod-fact}) is actually an expansion in powers of $v$. For
charmonium, $v^2\approx 0.3$; for bottomonium $v^2\approx 0.1$.
%
%

An important feature of the NRQCD factorization formalism is that
quarkonium decay and production occur through color-octet, as well as
color-singlet, $Q\overline Q$ states.  If one drops all of the
color-octet contributions, then the result is the color-singlet model
(CSM).  In contrast, NRQCD factorization is not a model.  It sometimes
is called, erroneously, ``the color-octet model,'' but it is, rather, a
consequence of QCD in the limit $m$,~$p_T\gg \Lambda_{\rm QCD}$.

A proof of NRQCD factorization would rely both on NRQCD itself and on the
machinery that is used to prove factorization of hard-scattering
processes in QCD. Although it is widely believed that the NRQCD
factorization formula (\ref{prod-fact}) can be established by standard
methods, no detailed proof exists in the literature. It is expected that
corrections to the factorization formula are of order $\Lambda_{\rm
QCD}^2/p_T^2$ for unpolarized cross sections and $\Lambda_{\rm QCD}/p_T$
for polarized cross sections \cite{qiu-sterman}.

\section{Some successes of the NRQCD factorization formalism}

Predictions of the NRQCD factorization formalism have been confirmed in 
a number of decay and production processes. In this talk, I will focus 
on quarkonium production.

\subsection{Quarkonium production at the tevatron}

\begin{figure}
\centerline{\includegraphics[width=8.2cm]{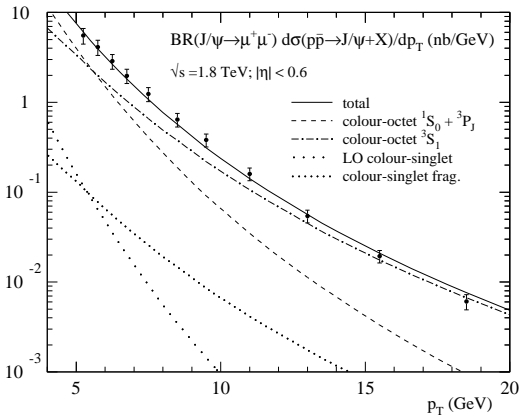}}
\caption{The $J/\psi$ cross section as a function of $p_T$. The data
points are from the CDF measurement \cite{Abe:1997jz}. The solid curve
is the NRQCD factorization fit to the data given in
Ref.~\cite{Kramer:2001hh}. The other curves give various contributions
to the NRQCD factorization fit. From
Ref.~\cite{Kramer:2001hh}.\label{fig:tevatron-j-psi}}
\end{figure}

The CDF data for $J/\psi$ production at the Tevatron are shown in
Fig.~\ref{fig:tevatron-j-psi}, along with a fit based on NRQCD 
factorization. 
As can be seen, the color-singlet contributions, whose normalizations
are known from decay processes, undershoot the data by more than an
order of magnitude. The color-octet contributions offer a possible
explanation of the data. However, it should be remembered that the
color-octet matrix elements are obtained by fitting to the Tevatron
data. A precise test of NRQCD factorization occurs only when one uses
the fitted values of the matrix elements to make predictions for other
processes. The shape of the data is consistent with NRQCD factorization,
although there is a good deal of freedom to change the theoretical shape
by adjusting the values of the color-octet matrix elements. The Tevatron
data for $\psi'$ and $\Upsilon$ production are also fit well by the
NRQCD factorization expressions.

\subsection{$\bm{\gamma \gamma\rightarrow J/\psi +X}$ at LEP}

\begin{figure}
\centerline{\includegraphics[width=6.3cm]{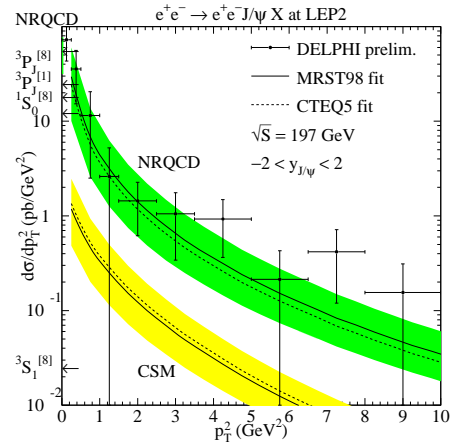}}
\caption{The cross section for ${\gamma \gamma\rightarrow J/\psi +X}$ at
LEP. The solid curves correspond to the MRST98LO \cite{Martin:1998sq}
parton distributions, and the dashed curves correspond to the CTEQ5L
\cite{CTEQ5} parton distributions. The upper set of curves is the
NRQCD factorization prediction, and the lower set of curves is the
color-singlet-model prediction. The data points are from the Delphi
measurement \cite{delphi}. From
Ref.~\cite{klasen-kniehl-mihaila-steinhauser}.\label{fig:delphi}}
\end{figure}

As can be seen in Fig.~\ref{fig:delphi}, in the case of $J/\psi$
production in $\gamma\gamma$ collisions at LEP, comparison of theory
\cite{klasen-kniehl-mihaila-steinhauser} with Delphi data clearly
favors NRQCD over the color-singlet model. 
Theoretical uncertainties arise mainly from uncertainties in the
color-octet matrix elements and from the choices of renormalization and
factorization scales. The latter uncertainties were estimated by varying
the scales up and down by a factor of two. The uncertainties in the
color-octet matrix elements are exacerbated by the fact that different
linear combinations of matrix elements appear in $J/\psi$ production at
LEP than in $J/\psi$ production at the Tevatron.

\subsection{Quarkonium production in DIS at HERA}

\begin{figure}
\centerline{\includegraphics[width=6.3cm]{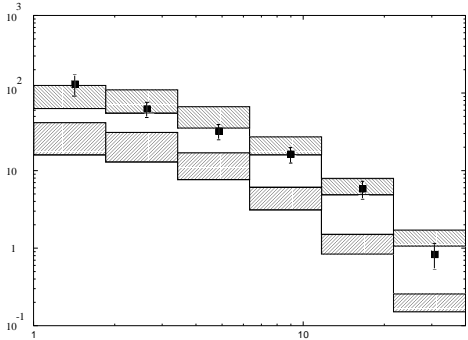}}
\caption{$J/\psi$ inclusive production in $ep$ DIS. The vertical axis is
$d\sigma/dp_T^2$ in units of pb$/$GeV$^2$; the horizontal axis is the
laboratory $p_T^2$ in units of GeV$^2$. The data points are from the H1
Collaboration \protect\cite{h1-dis-psi}. The upper histogram is the
leading-order NRQCD factorization prediction, and the lower histogram is
the CSM prediction. Theoretical error bands include estimates of
uncertainties from renormalization and factorization scales, parton
distributions, and color-octet matrix elements. From
Ref.~\cite{Kniehl:2001tk}. \label{fig:kz-PT2Lab}}
\end{figure}

The leading-order NRQCD factorization and CSM predictions of 
Ref.~\cite{Kniehl:2001tk} for the $J/\psi$ inclusive production cross
section $d\sigma/dp_T^2$ in $ep$ deep-inelastic scattering (DIS) are
shown in Fig.~\ref{fig:kz-PT2Lab}, along with the H1 data.
As can be seen, the data clearly favor the
NRQCD factorization prediction over the CSM prediction.  A similar
situation holds for $d\sigma/dQ^2$. Surprisingly,
the cross section $d\sigma/dz$, which is differential in the energy
fraction (inelasticity) $z$, is not fit well by either the
NRQCD factorization or CSM predictions. The former overshoots the data
at large $z$ and undershoots the data small $z$, while the latter
undershoots the data at all $z$. It is worth noting that the calculation
of Ref.~\cite{Kniehl:2001tk} disagrees with a number of previous results
\cite{Korner:1982fm,Guillet:1987xr,Merabet:sm,Krucker:1995uz,Yuan:2000cn},
which themselves are not fully consistent. Those discrepancies have not
yet been resolved completely.

\section{Some problematic comparisons with experiment}

There are several notable processes for which the comparison of
NRQCD factorization predictions with the data is less than satisfactory.

\subsection{Polarization of quarkonium at the Tevatron}

\begin{figure}
\centerline{\includegraphics[width=6.3cm]{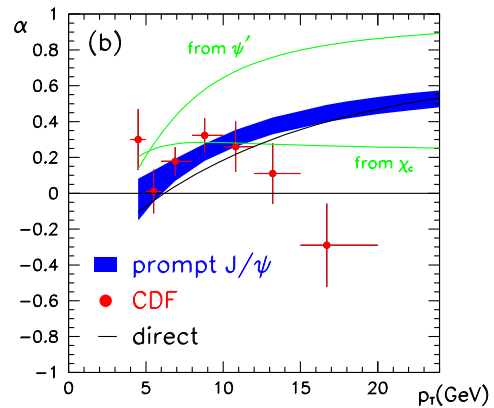}}
\caption{$J/\psi$ polarization at the Tevatron. The band is the
total NRQCD factorization prediction. The other curves give the
contributions from feeddown from higher charmonium states. The data
points are from the CDF measurement \cite{Affolder:2000nn}. From
Ref.~\cite{braaten-kniehl-lee-pol}.\label{fig:psi-pol}}
\end{figure}

The polarization of quarkonium produced at large $p_T$ at the Tevatron
provides a potentially definitive test of the color-octet mechanism.
Quarkonium production at large-$p_T$ ($p_T\agt 4m_c$ for the $J/\psi$) is
dominated by gluon fragmentation into the quarkonium through the
${}^3S_1$ color-octet matrix element. At large $p_T$, the fragmenting
gluon is nearly on its mass shell, and, so, is nearly transversely
polarized. According to the NRQCD velocity-scaling rules, spin-flip
interactions are suppressed relative to non-spin-flip interactions.
Therefore, it is expected that most of the gluon's polarization is
transferred to the $J/\psi$ \cite{cho-wise}. Radiative corrections and
color-singlet production dilute this polarization
\cite{beneke-rothstein-pol,Leibovich:1996pa,beneke-kramer-pol}.
Nevertheless, substantial polarization is expected at large $p_T$. In the
$J/\psi$ case, feeddown from the $\psi'$ and the $\chi_c$ states is
important and has now been taken into account
\cite{braaten-kniehl-lee-pol}. The NRQCD factorization prediction for
the $J/\psi$ polarization as a function of $p_T$ is shown, along with the
CDF data, in Fig.~\ref{fig:psi-pol}. 
The quantity $\alpha$ parametrizes the angular distribution of the decay
leptons in the $J/\psi$ rest frame: $d\sigma/d(\cos \theta) \propto
1+\alpha\cos^2\theta$. Here, $\theta$ is the angle between the
three-momentum of the positive lepton in the $J/\psi$ rest frame and the
boost vector from the $J/\psi$ rest frame to the CM frame of the
colliding hadrons. $\alpha=1$ corresponds to 100\% transverse
polarization;  $\alpha=-1$ corresponds to 100\% longitudinal
polarization. Polarization of produced $\psi'$ mesons is simpler
theoretically, since feeddown is not important, but, unfortunately, the
experimental statistics are much poorer.

As can been seen, the observed $J/\psi$ polarization is generally smaller
than the prediction and seems to trend in the wrong direction,
decreasing with increasing $p_T$.  However, the experimental error bars
are large, and only the last data point truly disagrees with the
prediction. Furthermore, there are large uncertainties in the
theoretical predictions. There are uncertainties in the NRQCD matrix
elements, which are reflected in the prediction band in
Fig.~\ref{fig:psi-pol}. There are uncertainties from uncalculated
contributions of higher order in $\alpha_s$, including effects from
multiple soft-gluon emissions and k-factors. Since the polarization
depends on a ratio of matrix elements, it probably is not strongly
affected by corrections to the matrix-elements fits from such
higher-order effects. Next-to-leading-order corrections have been
calculated for ${}^3S_1$ color-octet fragmentation
\cite{beneke-rothstein-pol,ma-frag,braaten-lee-frag}, which gives the
bulk of the polarization. Corrections to the non-fragmentation process
could conceivably increase the unpolarized contribution by a factor of two.
There are also large order-$v^2$ corrections to gluon fragmentation to
quarkonium \cite{Bodwin:2003wh}, but the principal effect of these is to
change the size of the corresponding color-octet matrix element in fits
to the Tevatron data, which does not affect the polarization prediction.
The large order-$v^2$ corrections do, however, raise questions as to the
convergence of the NRQCD $v$ expansion.

Existing calculations assume that 100\% of the $Q\overline Q$
polarization is transferred to the quarkonium. Spin-flip corrections are
suppressed only by $v^2$, not $v^4$, relative to the non-flip part
\cite{BBL}. It could happen that the spin-flip corrections are
anomalously large and depolarize the produced quarkonium. It has
also been suggested that the velocity-scaling rules may need to be
modified for the charmonium system
\cite{Brambilla:1999xf,Fleming:2000ib}. These issues should be resolved
by a lattice calculation that is in progress \cite{bodwin-lee-sinclair}.

\subsection{Inelastic photoproduction at HERA}

\begin{figure}
\centerline{\includegraphics[width=8.2cm]{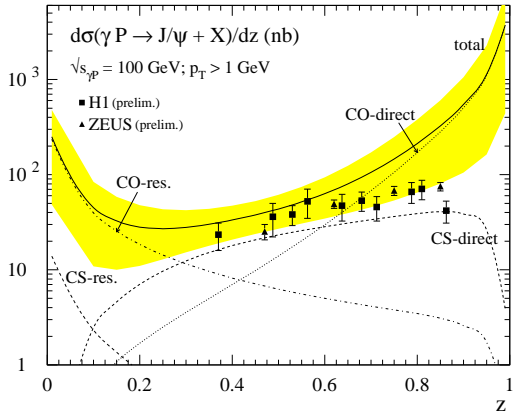}} 
\caption{The rate for inelastic quarkonium photo-production at HERA as a
function of the energy fraction $z$. The curves give total, direct, and
resolved color-octet (CO) and color-singlet (CS) contributions. The band
shows the uncertainty in the total contribution that arises from the
uncertainties in the color-octet matrix elements. The data points are
from the H1  and Zeus results of
Refs.~\cite{H1-photoproduction,Zeus-photoproduction}. The more recent H1
and Zeus results of Refs.~\cite{Adloff:2002ex,Chekanov:2002at} show a
similar behavior. From
Ref.~\cite{Kramer:2001hh}.\label{fig:photoproduction}}
\end{figure}

\begin{figure}
\centerline{\includegraphics[width=8.2cm]{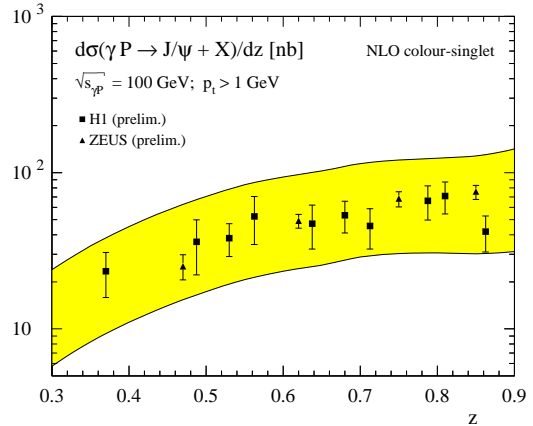}}
\caption{The rate for inelastic quarkonium photo-production at HERA as a
function of $z$, compared with the NLO color-singlet model. The band
shows the uncertainties that arise from $\alpha_s$ and $m$. The data
points are from the H1 and Zeus results of
Refs.~\cite{H1-photoproduction,Zeus-photoproduction}. From
Ref.~\cite{Kramer:2001hh}. \label{fig:photoproduction-CSNLO}}
\end{figure}

Theoretical calculations of the cross section for inelastic
photoproduction of quarkonium at HERA have been carried out in the
NRQCD factorization formalism by several groups
\cite{cacciari-kramer,Beneke:1998re,amundson-fleming-maksymyk,%
ko-lee-song,Godbole:1995ie,Kniehl:1997fv,Kniehl:1997gh}. The compilation 
of predictions from Ref.~\cite{Kramer:2001hh} and the H1 and Zeus
data are shown in Fig.~\ref{fig:photoproduction}, plotted as
function of the energy fraction $z$. 
As can be seen, the color-octet contribution is poorly determined, owing
to large uncertainties in the color-octet matrix elements. Even so,
there is little room for a color-octet contribution. Furthermore, as is
shown in Fig.~\ref{fig:photoproduction-CSNLO}, corrections of
next-to-leading order in $\alpha_s$ (NLO)
\cite{Kramer:1994zi,Kramer:1995nb} increase the color-singlet piece by
about a factor of two at large $z$ and are, by themselves, in good
agreement with the data.
The data shown in Figs.~\ref{fig:photoproduction} and
\ref{fig:photoproduction-CSNLO} are for the cut $p_T>1$~GeV. One
can question whether factorization is valid at such small values of
$p_T$. However, the data differential in $p_T$ are compatible with NLO
color-singlet production alone at large $p_T$ \cite{Kramer:2001hh}. It should
be noted, though, that there are large uncertainties in the NLO
color-singlet contribution, which arise primarily from uncertainties in
$m_c$ and $\alpha_s$. The true color-singlet contribution could be lower
than the central value by about a factor of two, leaving more room for a
color-octet contribution.

\begin{figure}
\centerline{\includegraphics[width=9.25cm]{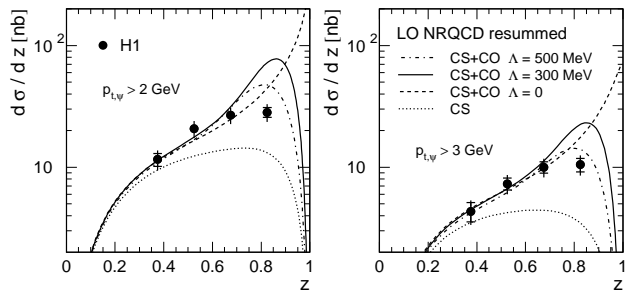}}
\caption{H1 
photoproduction data \cite{Adloff:2002ex} for $d\sigma/dz$ compared with
the leading-order color-octet (CO), color-singlet (CS), and total
(CO+CS) contributions for various values of the parameter $\Lambda$ in
the color-octet nonperturbative shape function. From
Ref.~\cite{Adloff:2002ex}. \label{fig:h1-shapefn}}
\end{figure}

Near $z=1$, the leading-order color-octet contribution grows rapidly, in
apparent disagreement with the data. However, in this, region soft-gluon
emission leads to large logarithms of $1-z$ and also to large
corrections of higher order in $v$, both of which must be resummed. The
resummation of the corrections of higher order in $v$ leads to a
nonperturbative ``shape function'' \cite{Beneke:1997qw}. Both the shape
function and the resummed logarithmic corrections significantly smear
out the color-octet contribution near $z=1$. The effects of a model
calculation of the shape function \cite{Beneke:1999gq} on the
leading-order NRQCD factorization predictions are shown in
Fig.~\ref{fig:h1-shapefn}, along with the H1 data.
The Zeus data \cite{Chekanov:2002at} show a similar behavior.
As can be seen, the inclusion of shape-function effects may lead to a
considerable improvement in the agreement of the NRQCD factorization
predictions with the data. The resummation of logarithms is expected to
have a comparable effect. (Note that the higher-$p_T$ data shown in
Fig.~\ref{fig:h1-shapefn} are more compatible with a color-octet
contribution than the data shown in Fig.~\ref{fig:photoproduction}).
Effects from resummation of logarithms of $1-z$ and model shape
functions have been calculated for the process $e^+e^-\to J/\psi+X$
\cite{Fleming:2003gt}. For shape functions that satisfy the
velocity-scaling rules, these effects are comparable in size. It may be
possible to use this resummed theoretical prediction to extract the shape
function from the Belle and BaBar data for  $e^+e^-\to J/\psi+X$ and
then use it to make firm predictions for $J/\psi$ photoproduction
near $z=1$.

\subsection{Double $\bm{c\overline{c}}$ production at Belle}

For the exclusive double charmonium process $e^+e^-\to J/\psi+\eta_c$,
the Belle Collaboration measures a cross section times a branching ratio
into at least four charged tracks of $46\pm 6^{+7}_{-9}~\hbox{fb}$
\cite{belle-eps2003}. In contrast, leading-order calculations predict a
cross section of $2.31\pm 1.09~\hbox{fb}$ 
\cite{Braaten:2002fi,Liu:2002wq, brodsky-ji-lee}. 
There are some uncertainties from uncalculated
corrections of higher-order in $\alpha_s$ and $v$ and from NRQCD matrix
elements. However, because this is an exclusive process, only
color-singlet matrix elements enter, and these are fairly well
determined from the decays $J/\psi\to e^+e^-$ and $\eta_c\to
\gamma\gamma$.

Since the Belle mass resolution is 110 MeV but the $J/\psi$-$\eta_c$
mass difference is only 120~MeV, it has been suggested that some of the
$J/\psi+\eta_c$ data sample may consist of $J/\psi+J/\psi$ events
\cite{Bodwin:2002fk,Bodwin:2002kk}. The state $J/\psi+J/\psi$ has
charge-parity $C=+1$, and consequently, is produced in a two-photon
process, whose rate is suppressed by a factor $(\alpha/\alpha_s)^2$
relative to the rate for $J/\psi+\eta_c$. However, as was pointed out in
Refs.~\cite{Bodwin:2002fk,Bodwin:2002kk}, the two-photon process
contains photon-fragmentation contributions that are enhanced by factors
$(E_{\rm beam}/2m_c)^4$ from photon propagators and $\log[8(E_{\rm
beam}/2m_c)^4]$ from a would-be collinear divergence. As a result, the
predicted cross section  $\sigma(e^+e^-\to J/\psi+J/\psi)=8.70\pm
2.94$~fb is larger than the predicted cross section $\sigma(e^+e^-\to
J/\psi+\eta_c)=2.31\pm 1.09$~fb, although corrections of higher order in
$\alpha$ and $v$ likely reduce the former prediction by about a factor
of three. These predictions spurred a re-analysis of the Belle data
\cite{Abe:2003ja}, with the result that there is no significant
$J/\psi+J/\psi$ signal observed [$\sigma(e^+e^-\to
J/\psi+J/\psi)<7~\hbox{fb}$].

There are also Belle results on inclusive double-charmonium production.
For the ratio $R_{J/\psi}=\sigma(e^+e^-\to J/\psi+c\overline c)
/\sigma(e^+e^-\to J/\psi+X)$, the most recent Belle analysis yields
$R_{J/\psi}=0.82\pm 0.15\pm 0.14$, with $R_{J/\psi}>0.48~\hbox{(90\%
confidence level)}$ \cite{belle-eps2003}. Predictions based on NRQCD
factorization \cite{Cho:1996cg,Baek:1996kq,Yuan:1996ep} give
$R_{J/\psi}\approx 0.1$. The  measured and predicted $J/\psi+c\overline
c$ absolute cross sections also disagree by almost an order of
magnitude, with the Belle result \cite{Abe:2002rb} being about
0.6--1.1~pb and the prediction \cite{Cho:1996cg,Baek:1996kq,Yuan:1996ep}
being about 0.10--0.15 pb. This prediction is based only on the
color-singlet contribution. However, corrections of higher order in $v$,
including color-octet contributions are not expected to be large.
Neither are corrections of higher order in $\alpha_s$.

The discrepancies in the double $c\overline c$ inclusive and exclusive
cross sections are among the largest in the standard model. Theory and
experiment differ by about an order of magnitude---a discrepancy which
is larger than any known QCD k-factor. It is important to recognize
that these discrepancies are problems not just for NRQCD factorization,
but for perturbative QCD (pQCD) in general. In the case of the cross
section for $e^+e^-\to J/\psi+\eta_c$, one obtains exactly the same
predictions in the NRQCD factorization \cite{Braaten:2002fi,Liu:2002wq}
and light-front-QCD \cite{brodsky-ji-lee} formalisms. With regard to the
fraction of $J/\psi+X$ events that are $J/\psi+c\overline c$, it is
difficult to see how any perturbative calculation could give a value as
large as 80\%. The color-evaporation model, for example, proceeds
through the same Feynman diagrams as the NRQCD factorization
calculation, differing only in the specific treatment of the evolution
of the $c\overline c$ pair into quarkonium, and would, therefore, be
expected to give a prediction that is not too different from that of
NRQCD factorization.

Clearly, it is very important to have independent checks of the Belle
double-charmonium results, such as could be provided by the BaBar
collaboration. If the Belle results are confirmed, then we would be
forced to entertain some unorthodox possibilities: there are new
charmonium production mechanisms within the standard model that have not
yet been recognized, pQCD is inapplicable to double-charmonium
production, or physics beyond the standard model plays an important
r\^ole. It would be very surprising, however, if the last possibility
could manifest itself at such low energies.

\section{Summary}

The NRQCD factorization approach provides a systematic method for
calculating quarkonium decay and production rates a double expansion in
powers of $\alpha_s$ and $v$. Calculation of production rates also
relies upon hard-scattering factorization, for which corrections are
suppressed by powers of $\Lambda_{\rm QCD}/p_T$. NRQCD factorization has
enjoyed a number of successes, for example, in quarkonium production at
the Tevatron, $\gamma\gamma\rightarrow J/\psi +X$ at LEP, and quarkonium
production in DIS at HERA. Other experimental tests are, so far, more
problematic. These include quarkonium polarization at the Tevatron,
inelastic quarkonium photoproduction at HERA,  and double $c\overline c$
production at Belle. The Belle double $c\overline c$ production results
present a severe challenge to pQCD. It would be very useful for the
BaBar collaboration to check these results. In other cases, inclusion of
corrections of higher order in $\alpha_s$ and $v$ and resummation of
soft-gluon effects and endpoint corrections of higher order in $v$
should help to achieve agreement between theory and experiment. More
precise theoretical predictions are hampered by uncertainties in the
NRQCD matrix elements. Lattice calculations can help to pin down the
decay matrix elements, but it is not yet known how to formulate the
calculation of production matrix elements on the lattice. This is an
exciting time for heavy-quarkonium physics, with a great deal of
experimental and theoretical activity in quarkonium decay and
spectroscopy, as well as production, and we can expect to see continuing
progress on the many challenging problems that remain.



\begin{thebibliography}{}

\bibitem{BBL}
G.~T.~Bodwin, E.~Braaten, and G.~P.~Lepage,
Phys.\ Rev.\ D \textbf{51}, 1125 (1995) [arXiv:hep-ph/9407339];
\textbf{55}, 5855(E) (1997).

\bibitem{Bodwin:1996tg}
G.~T.~Bodwin, D.~K.~Sinclair, and S.~Kim,
Phys.\ Rev.\ Lett.\  {\bf 77}, 2376 (1996) 
[arXiv:hep-lat/9605023].

\bibitem{Bodwin:2001mk}
G.~T.~Bodwin, D.~K.~Sinclair, and S.~Kim,
Phys.\ Rev.\ D {\bf 65}, 054504 (2002)
[arXiv:hep-lat/0107011].

\bibitem{qiu-sterman}
J.-w.~Qiu and G.~Sterman, private communication.

\bibitem{Abe:1997jz}
F.~Abe {\it et al.}  [CDF Collaboration],                                
Phys.\ Rev.\ Lett.\  {\bf 79}, 572 (1997). 

\bibitem{Kramer:2001hh}
M.~Kr\"amer,
Prog.\ Part.\ Nucl.\ Phys.\  {\bf 47}, 141 (2001)
[arXiv:hep-ph/0106120].

\bibitem{Martin:1998sq}
A.~D.~Martin, R.~G.~Roberts, W.~J.~Stirling, and R.~S.~Thorne,           
Eur.\ Phys.\ J.\ C {\bf 4}, 463 (1998) 
[arXiv:hep-ph/9803445].                                                  
   
\bibitem{CTEQ5}
H.~L.~Lai {\it et al.}  [CTEQ Collaboration],
Eur.\ Phys.\ J.\ C {\bf 12}, 375 (2000) 
[arXiv:hep-ph/9903282].

\bibitem{delphi}
S.~Todorova-Nova,                                                       
in {\it Proceedings of the XXXI International Symposium on Multiparticle
Dynamics}, Datong, China, 1--7 September, 2001 (World Scientific,       
Singapore, to appear) arXiv:hep-ph/0112050;                             
M.~Chapkin, talk presented at {\it 7th International Workshop on Meson
Production, Properties and Interaction (Meson 2002)}, Krakow, Poland,
May 24--28, 2002, unpublished.

\bibitem{klasen-kniehl-mihaila-steinhauser}
M.~Klasen, B.~A.~Kniehl, L.~N.~Mihaila, and M.~Steinhauser,             
Phys.\ Rev.\ Lett.\  {\bf 89}, 032001 (2002) 
[arXiv:hep-ph/0112259].                          

\bibitem{Kniehl:2001tk}
B.~A.~Kniehl and L.~Zwirner,                                            
Nucl.\ Phys.\ {\bf B621}, 337 (2002) 
[arXiv:hep-ph/0112199].                                              

\bibitem{h1-dis-psi}                                           
A.~Meyer,          
DESY-THESIS-1998-012;                                                 
A.~Meyer (unpublished); S.~Mohrdieck (unpublished).

\bibitem{Korner:1982fm}
J.~G.~Korner, J.~Cleymans, M.~Kuroda, and G.~J.~Gounaris,
Phys.\ Lett.\ B {\bf 114}, 195 (1982).                                    
                   
\bibitem{Guillet:1987xr}
J.~P.~Guillet,
Z.\ Phys.\ C {\bf 39}, 75 (1988).

\bibitem{Merabet:sm}
H.~Merabet, J.~F.~Mathiot, and R.~Mendez-Galain,
Z.\ Phys.\ C {\bf 62}, 639 (1994).

\bibitem{Krucker:1995uz}
D.~Kr\"ucker, Ph.D.\ Thesis, RWTH Aachen, 1995.

\bibitem{Yuan:2000cn}
F.~Yuan and K.~T.~Chao,
Phys.\ Rev.\ D {\bf 63}, 034017 (2001) 
[arXiv:hep-ph/0008301].

\bibitem{cho-wise}
P.~L.~Cho and M.~B.~Wise,                                               
Phys.\ Lett.\ B {\bf 346}, 129 (1995) 
[arXiv:hep-ph/9411303].                                                 

\bibitem{beneke-rothstein-pol}
M.~Beneke and I.~Z.~Rothstein,                                     
Phys.\ Lett.\ B {\bf 372}, 157 (1996) 
[Erratum-ibid.\ B {\bf 389}, 769 (1996)]                           
[arXiv:hep-ph/9509375].                                                 

\bibitem{Leibovich:1996pa}                                        
A.~K.~Leibovich,                                                        
Phys.\ Rev.\ D {\bf 56}, 4412 (1997) 
[arXiv:hep-ph/9610381].
         
\bibitem{beneke-kramer-pol}
M.~Beneke and M.~Kr\"amer,                                
Phys.\ Rev.\ D {\bf 55}, 5269 (1997) 
[arXiv:hep-ph/9611218].            

\bibitem{braaten-kniehl-lee-pol}
E.~Braaten, B.~A.~Kniehl, and J.~Lee,                                    
Phys.\ Rev.\ D {\bf 62}, 094005 (2000) 
[arXiv:hep-ph/9911436].                                                  

\bibitem{Affolder:2000nn}                                          
T.~Affolder {\it et al.}  [CDF Collaboration],
Phys.\ Rev.\ Lett.\  {\bf 86}, 3963 (2001).                         

\bibitem{ma-frag}
J.~P.~Ma,                                                               
Nucl.\ Phys.\ {\bf B447}, 405 (1995) 
[arXiv:hep-ph/9503346].                                             
                                                                         
\bibitem{braaten-lee-frag}
E.~Braaten and J.~Lee,
Nucl.\ Phys.\ B {\bf 586}, 427 (2000)
[arXiv:hep-ph/0004228].

\bibitem{Bodwin:2003wh}
G.~T.~Bodwin and J.~Lee,
arXiv:hep-ph/0308016.

\bibitem{Brambilla:1999xf}
N.~Brambilla, A.~Pineda, J.~Soto and A.~Vairo,
Nucl.\ Phys.\ B {\bf 566}, 275 (2000) 
[arXiv:hep-ph/9907240].

\bibitem{Fleming:2000ib}
S.~Fleming, I.~Z.~Rothstein and A.~K.~Leibovich,
Phys.\ Rev.\ D {\bf 64}, 036002 (2001) 
[arXiv:hep-ph/0012062].

\bibitem{bodwin-lee-sinclair}
G.~T.~Bodwin, J.~Lee, D.~Sinclair, work in progress.

\bibitem{cacciari-kramer}
M.~Cacciari and M.~Kr\"amer,
Phys.\ Rev.\ Lett.\  {\bf 76}, 4128 (1996) 
[arXiv:hep-ph/9601276].

\bibitem{Beneke:1998re}
M.~Beneke, M.~Kr\"amer, and M.~Vanttinen,
Phys.\ Rev.\ D {\bf 57}, 4258 (1998) 
[arXiv:hep-ph/9709376].

\bibitem{amundson-fleming-maksymyk}
J.~Amundson, S.~Fleming, and I.~Maksymyk,
Phys.\ Rev.\ D {\bf 56}, 5844 (1997) 
[arXiv:hep-ph/9601298].

\bibitem{ko-lee-song}
P.~Ko, J.~Lee and H.~S.~Song,
Phys.\ Rev.\ D {\bf 54}, 4312 (1996);
{\bf 60}, 119902(E) (1999)]
[arXiv:hep-ph/9602223].

\bibitem{Kniehl:1997fv}
B.~A.~Kniehl and G.~Kramer,
Phys.\ Lett.\ B {\bf 413}, 416 (1997) 
[arXiv:hep-ph/9703280].

\bibitem{Kniehl:1997gh}
B.~A.~Kniehl and G.~Kramer,
Phys.\ Rev.\ D {\bf 56}, 5820 (1997) 
[arXiv:hep-ph/9706369].


\bibitem{Godbole:1995ie}
R.~M.~Godbole, D.~P.~Roy, and K.~Sridhar,
Phys.\ Lett.\ B {\bf 373}, 328 (1996) 
[arXiv:hep-ph/9511433].

\bibitem{H1-photoproduction}
H1 Collaboration, Contributed paper 157aj, International Europhysics
Conference on High Energy Physics (EPS99), Tampere, Finland, 1999.

\bibitem{Zeus-photoproduction}
Zeus Collaboration, Contributed Paper 851, International Conference on
High Energy Physics (ICHEP2000), Osaka, Japan, 2000.

\bibitem{Adloff:2002ex}
C.~Adloff {\it et al.}  [H1 Collaboration],
Eur.\ Phys.\ J.\ C {\bf 25}, 25 (2002) 
[arXiv:hep-ex/0205064].

\bibitem{Chekanov:2002at}
S.~Chekanov {\it et al.}  [ZEUS Collaboration],
Eur.\ Phys.\ J.\ C {\bf 27}, 173 (2003) 
[arXiv:hep-ex/0211011].

\bibitem{Kramer:1994zi}
M.~Kr\"amer, J.~Zunft, J.~Steegborn, and P.~M.~Zerwas,                   
Phys.\ Lett.\ B {\bf 348}, 657 (1995) 
[arXiv:hep-ph/9411372].                                   
                        
\bibitem{Kramer:1995nb}
M.~Kr\"amer,
Nucl.\ Phys.\ B {\bf 459}, 3 (1996) 
[arXiv:hep-ph/9508409].

\bibitem{Beneke:1997qw}
M.~Beneke, I.~Z.~Rothstein, and M.~B.~Wise,                       
Phys.\ Lett.\ B {\bf 408}, 373 (1997) 
[arXiv:hep-ph/9705286].                                                

\bibitem{Beneke:1999gq}                                             
M.~Beneke, G.~A.~Schuler, and S.~Wolf,
Phys.\ Rev.\ D {\bf 62}, 034004 (2000) 
[arXiv:hep-ph/0001062].                                             

\bibitem{Fleming:2003gt}
S.~Fleming, A.~K.~Leibovich, and T.~Mehen,
arXiv:hep-ph/0306139.

\bibitem{belle-eps2003}              
K.~Abe {\it et al.}  [Belle Collaboration], BELLE-CONF-0331, contributed 
paper, International Europhysics Conference on High Energy Physics
(EPS 2003), Aachen, Germany, 2003.

\bibitem{Braaten:2002fi}
E.~Braaten and J.~Lee,
Phys.\ Rev.\ D {\bf 67}, 054007 (2003) 
[arXiv:hep-ph/0211085].

\bibitem{Liu:2002wq}
K.~Y.~Liu, Z.~G.~He, and K.~T.~Chao,
Phys.\ Lett.\ B {\bf 557}, 45 (2003) 
[arXiv:hep-ph/0211181].

\bibitem{brodsky-ji-lee}
S.~J.~Brodsky, C.-R.~Ji, and J.~Lee, private communication.

\bibitem{Bodwin:2002fk}
G.~T.~Bodwin, J.~Lee, and E.~Braaten,
Phys.\ Rev.\ Lett.\  {\bf 90}, 162001 (2003) 
[arXiv:hep-ph/0212181].

\bibitem{Bodwin:2002kk}
G.~T.~Bodwin, J.~Lee, and E.~Braaten,
Phys.\ Rev.\ D {\bf 67}, 054023 (2003) 
[arXiv:hep-ph/0212352].

\bibitem{Abe:2003ja}
K.~Abe {\it et al.}  [Belle Collaboration],
arXiv:hep-ex/0306015.

\bibitem{Cho:1996cg}
P.~L.~Cho and A.~K.~Leibovich,
Phys.\ Rev.\ D {\bf 54}, 6690 (1996)
[arXiv:hep-ph/9606229].

\bibitem{Baek:1996kq}
S.~Baek, P.~Ko, J.~Lee, and H.~S.~Song,
Phys.\ Lett.\ B {\bf 389}, 609 (1996) 
[arXiv:hep-ph/9607236].

\bibitem{Yuan:1996ep}
F.~Yuan, C.~F.~Qiao, and K.~T.~Chao,
Phys.\ Rev.\ D {\bf 56}, 321 (1997) 
[arXiv:hep-ph/9703438].

\bibitem{Abe:2002rb}
K.~Abe {\it et al.}  [Belle Collaboration],
Phys.\ Rev.\ Lett.\  {\bf 89}, 142001 (2002)
[arXiv:hep-ex/0205104].

\end{thebibliography}
\end{document}